# Entropy and Information is Transferred from Peripherical Sites to the Catalytic Sites of Enzymes

Germán Miño-Galaz[1*], Juan Pablo Peña[2], Javier Patiño Baez[3], Nicolas Miño-Berdú[4] and José González Suárez[3]

[1] Facultad de Ingeniería Ciencia y Tecnología, Universidad Bernardo O'Higgins, Av. Viel 1497, Santiago, Región Metropolitana, Chile

[2] Departamento de Matematicas, Facultad de Ciencias Exactas, Universidad Andres Bello, República 498, Santiago, Chile

[3] Departamento de Ciencias Físicas, Facultad de Ciencias Exactas, Universidad Andres Bello, República 498, Santiago, Chile

[4] Departamento de Matemática y Ciencia de la Computación, Universidad de Santiago de Chile, Santiago, Chile

## Abstract

This research reports the entropy and information transfer throughout seven different enzymatic systems, namely, TIM-Barrel, Human Lysozyme, Ribonuclease A1, Pepsin , β-lactamase, Human Glucokinase and Carbonic anhydrase II. A general trend is detected: entropy and information is transported form the peripherical regions towards the catalytic site of the analyzed enzymatic systems.

## Graphical Abstract

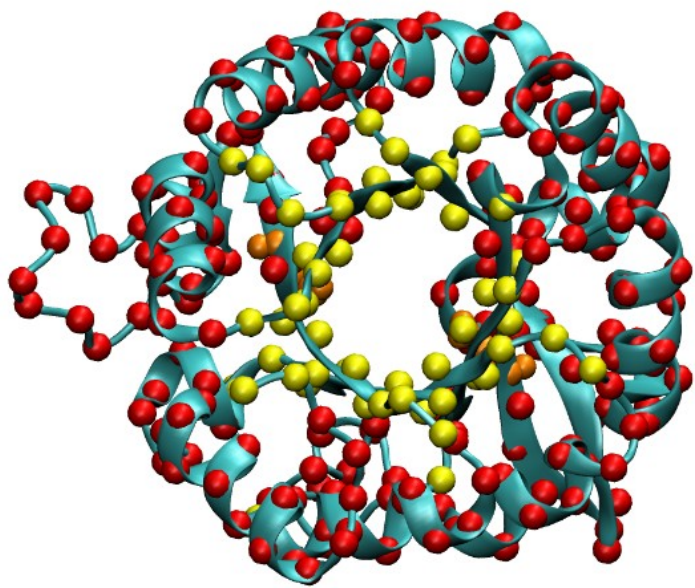

## Introduction

Information flux throughout the three-dimensional structure of proteins and enzymes is necessary for the modulation of their biological activity. This information is processed by the structural and topological features of the protein itself and may lead to the activation or deactivation of the biological activity of a given protein or enzyme. This phenomenon is it know in the field of biochemistry as allosteric communication. The word allosteric comes from the composite Greek expression in which "allos" means different and "steric" means place, so the modulation and communication comes from "a different place" of the protein. Allosteric communication requires at least two sites that interact with each other. At one hand, there is the allosteric site where an action is performed by a molecule or ligand acting to generate an input or cause; on the other hand, there is an effector site that executes a response. The separation between the allosteric and the effector sites spans from short to distances reaching hundreds of Å´s along the three-dimensional structure of a given protein. The physical nature of the signal and of the information that mediates this kind of molecular communication points to an entropy mediated phenomenon (Hacisuleyman 2017). It has been established that this entropy, in the sense of Shanon´s entropy, is transported throughout protein structure by a vibrational mediated mechanism and operates in the following way: if the time fluctuation of a given residue drives the future time fluctuation of another residue a source → sink relationship is established. Due to the intricate three-dimensional structures of a folded protein an asymmetrical information flow may holds. This information transfer throughout the three-dimensional structure of proteins leads to the modulation of the of the effector site, which may consist in a chemically active machinery or a docking site for another substrate. For example, negative modulation is observed when cAMP molecules bind to the catabolite activator protein (CAP). There, the binding of the initial cAMP molecule to CAP makes the binding of a second cAMP more difficult. (Popovych, 2006). Several methods, experimental and computational, are used to study allosteric communication. (Miño-Galaz 2015, Liu 2016, Chen 2022, Morales-Pastor 2022, Bernetti 2024, Nerin-Fonz 2024, Wagner 2016, Verkhivker 2020). One method, the coarse grain dynamic Gaussian Network Model (dGNM), offers the advantage of a very low computational cost with the ability to measure relevant and precise information about intramolecular entropy transfer (Haliloglu 1997, Hacisuleyman 2017, Erkip, 2004). In dGNM the atomic details are erased and only the information of alpha carbon atoms ($C^\alpha$) of the protein are preserved (Figure 1). This network is allowed to interact by a distance dependent single harmonic parameter force field. This methodology have been validated for the prediction of allosteric features of proteins systems (Erman 2017, Haliloglu 2022, Erman 2023). By measuring the spontaneous fluctuations of $C^\alpha$ about their average position, it is possible to track entropy transfer between pairs of positions in a protein system. Using the connectivity matrix between $C^\alpha$ of the protein (also named the graph Laplacian $\Gamma$), dGNM algorithm calculates time delayed correlations useful determine entropy and information transfer between pairs of $C^\alpha$. The three-dimensional fold of a protein determines which residues are in contact, and this connectivity pattern constrains the fluctuations of each residue and shapes the pathways through which information can propagate (Erman 2017, 2023).

A complementary perspective emerges from the structural and topological analysis of the protein topological contact network. The intrinsic correlation structure obtainable by dGNM is not a dynamical accident but is rooted in the topology of the contact network: the graph Laplacian $\Gamma$ encodes this topology directly. Its pseudoinverse $\Gamma^{-1}$ determines the covariance of residue fluctuations, and since fluctuations are well approximated by a multivariate Gaussian distribution (Erman 2023), the mutual information between any pair of residues can be expressed in closed form in terms of $\Gamma^{-1}$ alone. This allows us to define an information centrality $IC_i$ for each residue that quantifies its structural capacity of a given $C^\alpha$ to act as an information hub—a purely topological measure that requires no dynamical simulation. This fact offers a secondary and simpler way, used here, to measure information flow: information centrality (IC), which is defined as the normalized information flux that a given residue or $C^\alpha$ manages.

In this research, we study both dGNM and IC to track the intramolecular entropy and information transfer in seven different enzymatic systems by directly determining the transfer capability per $C^\alpha$. This capability is mapped upon the three-dimensional each protein structure revealing a the general directionality of the signal transfer process. The analysis was performed in the TIM-Barrel (PDBid 5TQL), Human Lysozyme (PDB iD:1RE2), Ribonuclease A1 (PDB id: 1RUV), Pepsin (PDB id: 4PEP), β-lactamase (PDB id:1BTL), Human Glucokinase (PDB id: 1V4S), Carbonic anhydrase II (PDB id: 1HEA). We found an entropy and information gradient in which the flux occurs from peripherical sites to the catalytic sites of each of those enzymes. Also is found that both the dGNM and IC gives the same essential results. In this report we make a general inspection of how information is propagated throughout enzymatic protein three-dimentional structure, more than to make a specific allocation of allosteric sites or allosteric pathways.

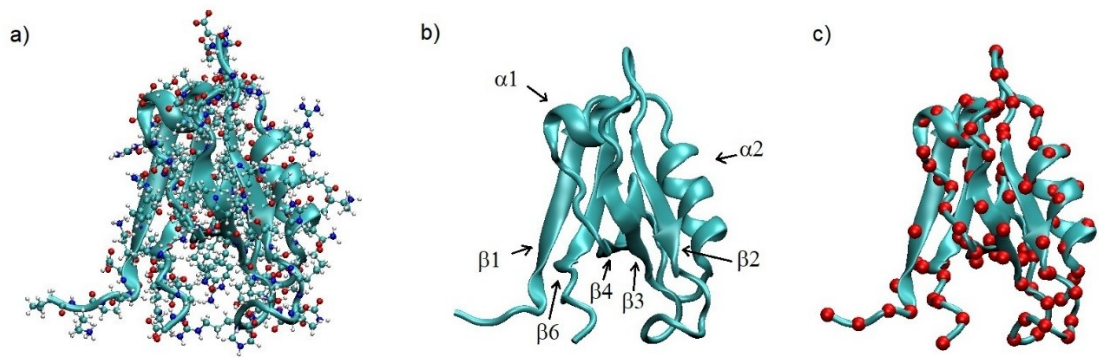

Figure 1. Reduction from all atom representation to the $C^\alpha$ only model using the PDZ-2 protein as example (not analyzed in this paper). Three representations are depicted a) all atoms superimposed with ribbon representation. b) ribbon representation depicting the labeling of the secondary structural elements, c) Ribbon representation plus the alpha carbon atoms ($C^\alpha$) representation shown in red spheres. The lone $C^\alpha$ coordinates are used as input of the dGNM and for IC methods.

## Theoretical methods and computational details.

The dGNM model for a folded protein assigns interaction of close residues by linear springs, thus the protein is analyzed in the context of a three-dimensional elastic network. The junctions or nodes are the alpha carbon atoms ($C^\alpha$) of the protein. The position of the i-th $C^\alpha$ is denoted by $R_i$ and its gaussian fluctuation around the equilibrium position is expressed as $\Delta R_i(t) = R_i - \overline{R}_i$ where $\overline{R}_i$ is the mean position. The dGNM method uses a connectivity matrix $\Gamma$ that is defined as:

$$\Gamma_{ij} = \left\{ \begin{array}{ll} -1 & \text{if } i \neq j \text{ and } R_{ij} \leq r_c, \\ 0 & \text{if } i \neq j \text{ and } R_{ij} > r_c, \\ -\sum_{i, i \neq j} \Gamma_{ij} & \text{if } i = j. \end{array} \right\}$$

Equation 1.

Here $R_{ij}$ is the distance between the $i^{th}$ and $j^{th}$ atoms, and $r_c$ is a cut-off distance of 7.0 Å (in our case). The diagonal elements are the negative sum of the off-diagonal elements of the $i^{th}$ column. So, $\Gamma$ is an n x n symmetric matrix, where n is the number of residues. $\Gamma$ is also known as local packing or coordination density matrix. The diagonalization of $\Gamma$ is expressed as $\Gamma^{-1} = U(\Lambda)^{-1} U^T$, where **U** is an orthogonal matrix with columns of the eigenvectors $u_i$ of the matrix $\Gamma$, and $\Lambda$ is a diagonal matrix composed by the eigenvalues $\lambda_i$ of $\Gamma$. It is possible to decompose $\Gamma^{-1}$ as the sum of contributions from individual modes as

$$\Gamma^{-1} = \sum_{k=2}^{n} \lambda_k^{-1} u_k u_k^T = \sum_{k=2}^{n} A^{(k)}$$

Equation 2.

Where $A_{ij}^{(k)}$ is an n x n matrix that describes the contribution of the k-th vibrational mode to the atomic fluctuations, thus *k=2*. The first eigenvalue of $\Gamma$ is zero is not included in the summation. The time delayed correlation between the $i^{th}$ and $j^{th}$ $C^{\alpha}$ atoms can be expressed as the sum of the contributions of individual modes as

$$\langle \Delta R_i(0) \Delta R_j(\tau) \rangle = \sum_k A_{ij}^{(k)} \exp\left(\frac{-\lambda_k \tau}{\tau_0}\right)$$

Equation 3.

Here $\tau$ is the time delay, that we selected for our simulations as 1.2 ps and $\tau_0$ is a fixed time of 6 ps, which is characteristic of the vibrational dynamics of all proteins in their folded state (Haliloglu 1997). Applying Schreiber's work in information transfer (Schreiber 2000), Erman describes a way to quantify the time delayed information transfer between pairs of residues in a protein (Hacisuleyman, 2017). So, information transfer depends on conditional probabilities evaluated for the fluctuations of residues i and j. The fluctuation of the two-vector positions $\Delta R_i(t)$ and $\Delta R_j(t)$ change rapidly with time. If the fluctuation of the two nodes is correlated, the knowledge of the fluctuation of the $i^{th}$ node (at t = 0) will decrease the uncertainty of the knowledge (at t = $\tau$) of the fluctuations of the $j^{th}$. The time dependence of two correlations is given by $\langle \Delta R_i(0) \Delta R_j(\tau) \rangle$, given by equation 3. If the asymmetric relation $\langle \Delta R_i(0) \Delta R_j(\tau) \rangle \neq \langle \Delta R_j(0) \Delta R_i(\tau) \rangle$ holds, a net information transfer from one residue to another is occurring. Thus, the entropy transfer $T_{i \rightarrow j}(\tau)$ from the trajectory $\Delta R_i(t)$ of residue i to the trajectory $\Delta R_j(t)$ of residue J is the amount of uncertainty loss in future values of $\Delta R_j(t+\tau)$ by knowing the past values of $\Delta R_j(t)$. The working equation, obtained by Erman, to determine the entropy transfer between two protein nodes is:

$$\begin{aligned}
T_{i \to j}(\tau) = &\frac{1}{2}\ln\left(\left(\sum_k A_{jj}^{(k)}\right)^2 - \left(\sum_k A_{jj}^{(k)} \exp\{-\lambda_k \tau/\tau_0\}\right)^2\right) \\
&-\frac{1}{2}\ln\left[\left(\sum_k A_{ii}^{(k)}\right)\left(\sum_k A_{jj}^{(k)}\right)^2\right. \\
&+2\left(\sum_k A_{ij}^{(k)}\right)\sum_k A_{jj}^{(k)} \exp\{-\lambda_k \tau/\tau_0\}\sum_k A_{ij}^{(k)} \exp\{-\lambda_k \tau/\tau_0\} \\
&-\left\{\left(\sum_k A_{ij}^{(k)} \exp\{-\lambda_k \tau/\tau_0\}\right)^2 + \left(\sum_k A_{ij}^{(k)}\right)^2\right\}\left(\sum_k A_{jj}^{(k)}\right) \\
&\left.-\left(\sum_k A_{jj}^{(k)} \exp\{-\lambda_k \tau/\tau_0\}\right)^2\left(\sum_k A_{ii}^{(k)}\right)\right] - \frac{1}{2}\ln\left[\left(\sum_k A_{jj}^{(k)}\right)\right] \\
&+\frac{1}{2}\ln\left(\left(\sum_k A_{ii}^{(k)}\right)\left(\sum_k A_{jj}^{(k)}\right) - \left(\sum_k A_{ij}^{(k)}\right)^2\right)
\end{aligned}$$

Equation 4.

With the elements of matrix A playing a central role in it. Again, $\tau$ is the time delay, and $\tau_0$ is a characteristic vibrational relaxation time as is shown for Equation 3. The solution leading to Equation 4 it is fully shown in Hacisuleyman 2017 and in its supplementary material. So, we calculated the information transfer based on

Equation 4 using a python-based algorithm. The calculation is done using only on frame of each structure. Each molecular structure was obtained from the Protein Data Bank (PDB). The algorithm extracts the coordinates of the $C^\alpha$ form the PDB file, constructs the $\Gamma$ matrix, determines the normal mode vector and eigenvalues to calculate the inverse $\Gamma^{-1}$. As $\Gamma^{-1} = \sum_{k=2}^{n} A^{(k)}$, its elements are used to calculate the residue-to-residue entropy transfer. Then we calculate the entropy transfer for one node to the rest of the protein using the summation

$$T_{i \to \oslash}(\tau) = \sum_j T_{i \to j}(\tau)$$

Equation 5.

Where $T_{i \to \oslash}(\tau)$ represents the total entropy that the i[th] element transfers to the rest of the nodes ($C^\alpha$) of the protein at the time delay $\tau$. Then, we map out the transfer capability of each node along the three-dimensional structure of the studied protein systems, and a general entropy transfer directionality is in this way be visualized.

Now we describe the IC methodology, and before introducing its formalism, it is useful to comment the simpler and more foundational model from which dGNM derives: the Gaussian Network Model (GNM) (Haliloglu 1997). While the dGNM described above captures the dynamics of residue fluctuations and their time-delayed correlations, the GNM provides a purely structural perspective on the same system. In similarity with dGNM the GNM, the folded protein is represented by the same Kirchhoff matrix $\Gamma$ (Equation 1), which is precisely the graph Laplacian, so the folded protein is represented as a graph $G=(V,E)$, where the nodes $V$ are the $C^\alpha$ atoms and the edges $E$ connect pairs of residues whose spatial distance is below the cutoff $r_c$. Therefore, GNM is intrinsically a topological model. Under the GNM, residue fluctuations follow a multivariate Gaussian distribution, and the covariance between the displacements of residues $i$ and $j$ is proportional to the $(i,j)$ element of the pseudoinverse of the Laplacian (Haliloglu 1997, Erkip 2004):

$$\langle \Delta R_i \cdot \Delta R_j \rangle \propto \Gamma^{-1}_{ij}$$

This is a key result: the dynamic correlation between two residues is entirely determined by the topology of the contact network, as encoded in $\Gamma^{-1}$. Residues that share a densely connected neighborhood will show correlated fluctuations even if they are spatially distant from one another. Since residue fluctuations in the GNM are Gaussian, the mutual information between the displacement of residue $i$ and that of residue $j$ can be expressed in closed form in terms of their correlation coefficient (Cover and Thomas, 2006). The normalized cross-correlation coefficient $C_{ij}$ is derived by Erman (2022) from the isotropic Gaussian approximation of the joint fluctuation probability. In that framework, $C_{ij}$ is expressed directly in terms of time-averaged displacement products:

$$C_{ij} = \frac{\langle \Delta R_i \cdot \Delta R_j \rangle}{\sqrt{\langle \Delta R_i^2 \rangle \langle \Delta R_j^2 \rangle}}$$

Equation 6.

where $\langle \Delta R_i \cdot \Delta R_j \rangle$ is the scalar correlation of the fluctuations of residues $i$ and $j$, and $\langle \Delta R_i^2 \rangle$ is the mean square displacement of residue $i$. This expression can be reinterpreted in the language of graph theory: in the GNM, these fluctuation averages are precisely the elements of the pseudoinverse of the graph Laplacian $\Gamma^{-1}$, so that $\langle \Delta R_i \cdot \Delta R_j \rangle \propto \Gamma^{-1}_{ij}$ and $\langle \Delta R_i^2 \rangle \propto \Gamma^{-1}_{ii}$ (Haliloglu 1997). The correlation $C_{ij}$ therefore becomes:

$$C_{ij} = \frac{\Gamma^{-1}_{ij}}{\sqrt{\Gamma^{-1}_{ii}}\sqrt{\Gamma^{-1}_{jj}}}$$

Equation 7.

a quantity that is entirely determined by the topology of the contact network, with no dynamical input required. For two jointly Gaussian variables, the mutual information then reduces to $-\frac{1}{2}\ln(1-C_{ij}^2)$ (Cover and Thomas, 2006), which connects the graph-topological structure of the protein directly to a quantitative measure of information sharing between residues.

The total information flux of residue i is obtained by summing its pairwise mutual information with all other residues in the protein:

$$F_i = \sum_{j \neq i} -\frac{1}{2}\ln(1-C_{ij}^2)$$

Equation 8.

$F_i$ captures how extensively a given residue participates in the global information network, irrespective of the direction of the transfer. The information centrality $IC_i$ is the normalised form of this flux, expressing each residue's contribution relative to the total information flow of the entire protein:

$$IC_i = \frac{F_i}{\sum_{k=1}^{N} F_k}$$

Equation 9.

where $N$ is the total number of residues. Importantly, $IC_i$ depends solely on $\Gamma^{-1}$ and therefore on the topology of the contact graph alone. This contrasts with the entropy transfer $T_{i \to j}(\tau)$ from the dGNM, which incorporates time-delayed correlations and captures the directional, dynamic propagation of information between pairs of residues (Hacisuleyman 2017). The two quantities are thus complementary: entropy transfer describes the dynamical process by which information flows through the network, while information centrality reflects the structural and topological capacity of each residue to act as an information hub. Both are derived from the same Laplacian operator, but they illuminate different facets of the same phenomenon. The convergence of both measures to the same qualitative picture—as shown in Figure 4—is not merely a numerical coincidence: it indicates that the allosteric architecture of enzymes is already encoded at the level of the contact topology. This justifies the structural and topological analysis of protein–enzyme molecules as a productive strategy for identifying the residues that govern information flow towards the catalytic site, even in the absence of detailed dynamical simulations.

**Results.**

We explored the distribution of the capability of a given $C^\alpha$ to propagate entropy and information in seven different enzymatic systems. The exploration is performed by plotting this property upon the three-dimensional structure of the protein structures. For this we use Equation 5, which is the summation of the entropy transfer from one node or $C^\alpha$ to the rest of the molecule. Then, we ranked the $C^\alpha$ according to their capability and inspected its allocation upon the three-dimensional structure. Specifically, we analyzed the TIM-Barrel protein HisF-C9S (PDBid 5TQL) which contain 249 amino-acids residues in its crystallographic structure. This class of biochemical structure is known as alpha/beta barrel. So, we pass its structure through the dGNM algorithm (using Equation 4) and generate the list of best/worst information donors (using Equation 5. The, we allocate them upon the three-dimensional structure of the barrel. The numerical result is available as supporting information and the visual depiction of it is shown in Figure 2. Figure 2c shows the cartoon backbone structure of the TIM-Barrel depicting in orange the residues Asp130 and Asp11 just as reference due they are part of the catalytic residues of this system (Beismann-Driemeyer 2001) (atoms of those catalytic residues are not part of the calculation and only $C^\alpha$ are part of the dGNM calculation). Figure 2a shows the 177 best (red spheres) and the 72 worst (yellow spheres) donors allocated upon the TIM-Barrel Structure. Figure 2d shows sagittal view with the allocation of the worst 77 information donors and the location of Asp130 and Asp11 residues. Here is observed the existence of an entropy gradient defined by the best/worst information donors, with the central region harboring the worst donors, while the peripherical region containing the best entropy donors. It's clear from the visualization that entropy propagates from the outer regions of this protein to the inner regions of it to the central region that contains the catalytic residues. Also, the three-dimensional projection also shows that the entropy is transferred mainly to the hydrophobic regions of the protein (Figure 2d). It is known that this TIM-Barrel structure has a lower N-terminal region of the β-sheet (ab loop region) that is dedicated to give stability to the protein, and a C-terminal region of the β-sheet (βα loop region) that is dedicated to hold the catalytic activity (Wierenga 2001). Our analysis of best/worst donor clearly shows that the entropy is transported very close to the region containing the catalytic residues. It is important to mention that the TIM-Barrel structure represents a ubiquitous scaffold that hosts at least 15 distinct enzyme families in a wide range of living organisms (Wierenga 2001). Also, the TIM-Barrel general organization is actively used for de novo enzymatic design and engineering (Romero-Romero 2021) and allosteric modulation has been reported for this kind of biological macrostructure (Gamiz-Arco 2021, Chan 2020).

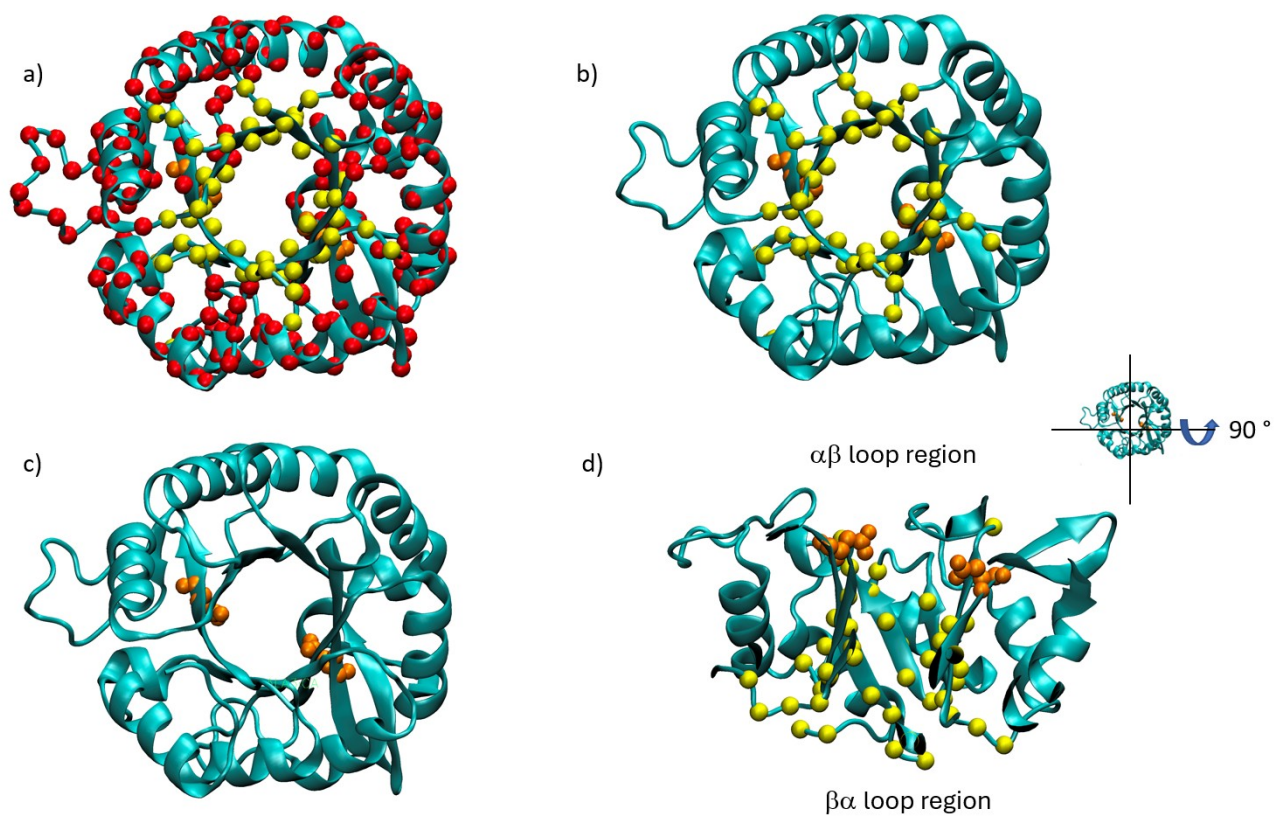

a) b) c) d) αβ loop region   90°

βα loop region

Figure 2. Allocation of the best/worst donors upon the 3-dimensional structure of the TIM-Barrel protein HisF-C9S (PDB id 5TQL). a) Cartoon diagram of the protein highlighting the catalytic residues Asp130 and Asp11 in orange (as reference) with 177 best (red spheres) and 70 worst (yellow spheres) entropy donors allocated upon the structure b) view of the catalytic residues Asp130 an Asp11 and the 56 worst (yellow spheres) allocated upon the structure c) view of the catalytic residues Asp130 an Asp11 upon the TIM-Barrel. d) Sagittal view of the TIM-barrel showing that allocation of the worst donors with respect to the catalytic residues. The worst donors are allocated in stabilization region ($\beta\alpha$ loop) next to the catalytic region ($\alpha\beta$ loop).

After this analysis, we evaluated of both entropy (using dGNM) and information flow (using IC) in other six monomeric enzymatic systems, namely Human Lysozyme (PDB iD:1RE2), Ribonuclease A1 (PDB id: 1RUV), Pepsin (PDB id: 4PEP), $\beta$-lactamase (PDB id:1BTL), Human Glucokinase (PDB id: 1V4S), Carbonic anhydrase II (PDB id: 1HEA). The result of this analysis is shown in Figure 3. For all the systems the catalytic residues are depicted in orange as reference for the catalytic site allocation. In general, there is an entropy gradient that is allocated going form peripherical residues, with the highest values, to the catalytic residues, with lowest values in all analyzed cases. This fact is revealed as general trend, in which information and entropy is transferred towards the catalytic region of all the analyzed enzymes. Next, we calculated the IC for each of the six enzymatic systems. The graphical results are presented in Figure 4. It is clear from the observation of Figure 4 that entropy transfer calculated by dGNM and information centrality calculated by IC give essentially the same results. The observed gradient is completely consistent with the general paradigm of allosteric communication, that is, that information should be transported from any other site to the biologically active site of enzymes.

I
a) 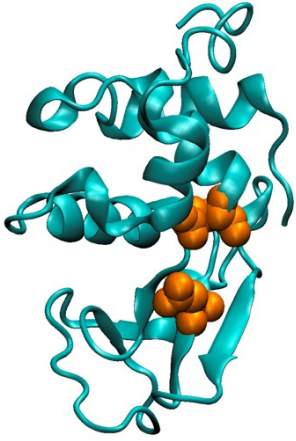 b) 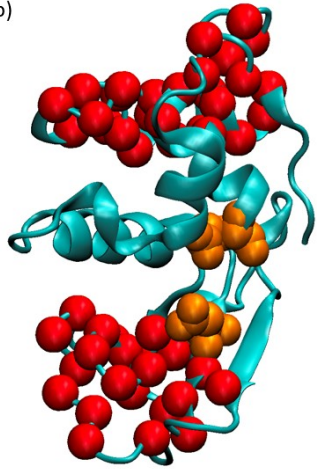 c) 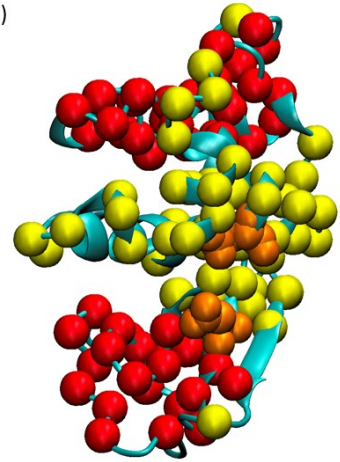

II
a) 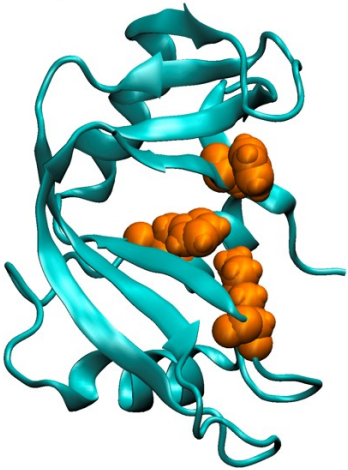 b) 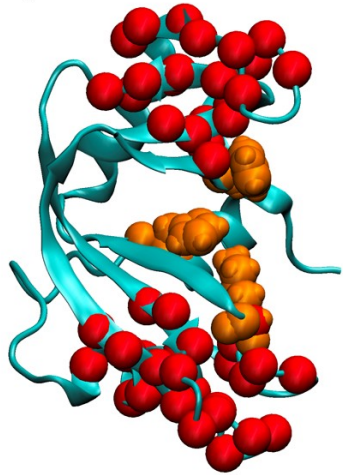 c) 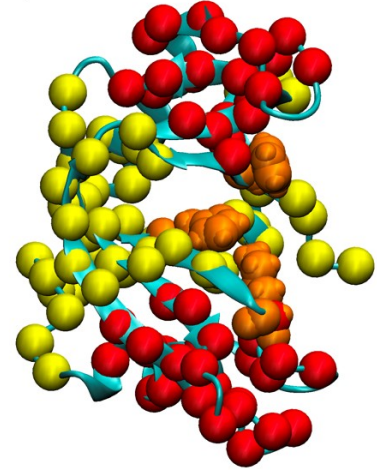

III
a) 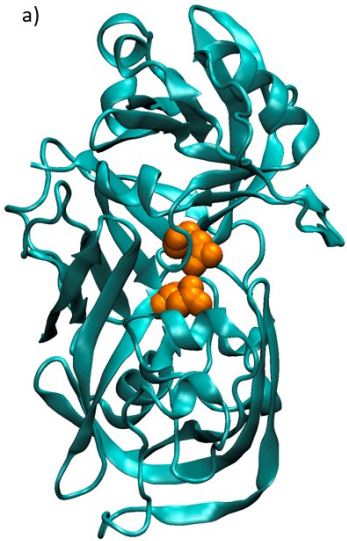 b) 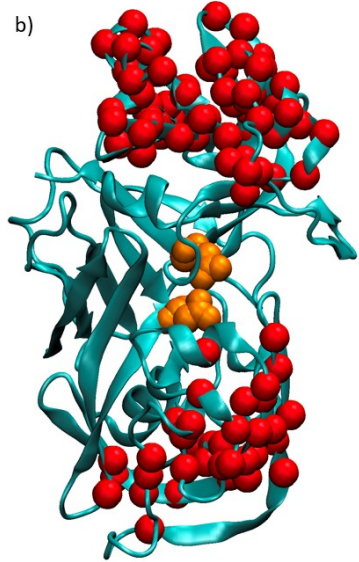 c) 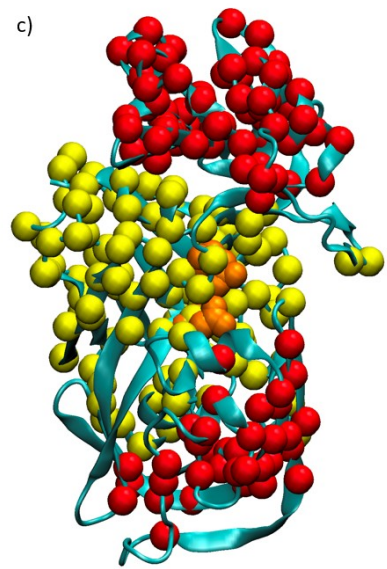

IV
a) 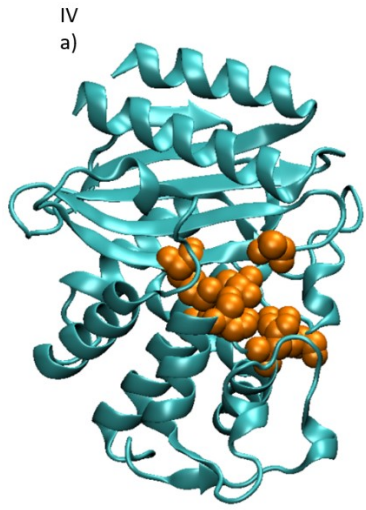 b) 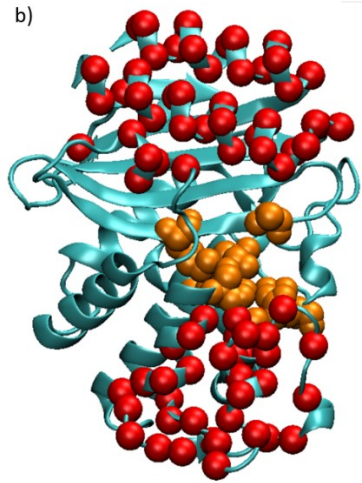 c) 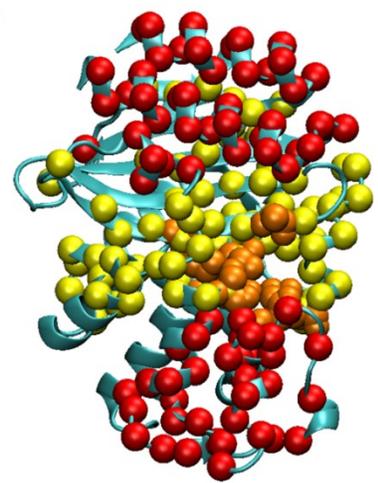

V
a) 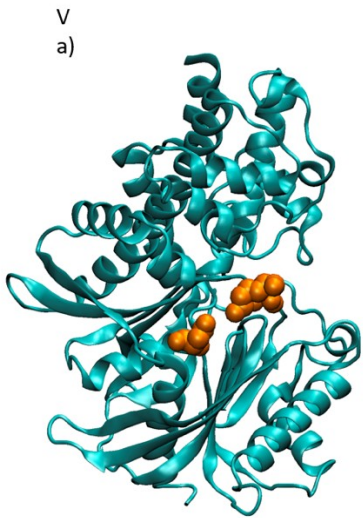 b) 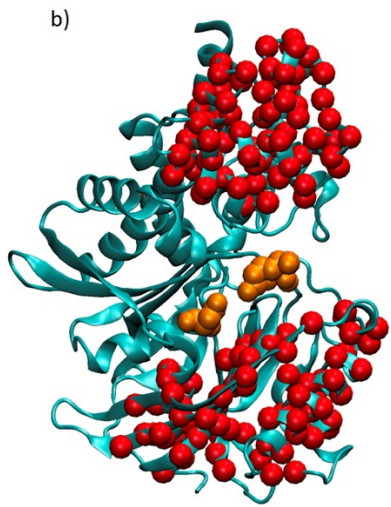 c) 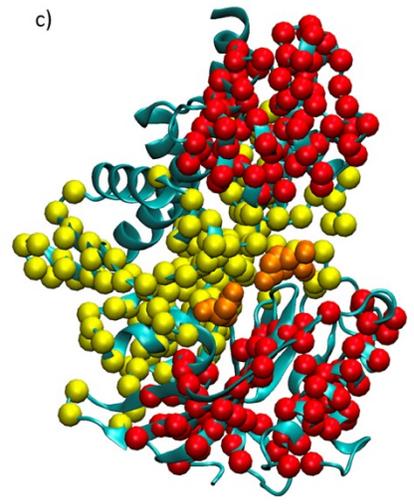

VI
a) 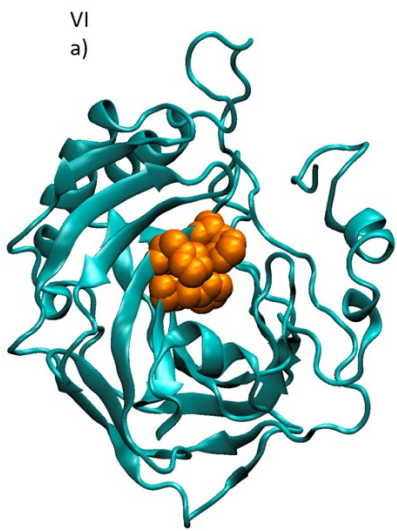 b) 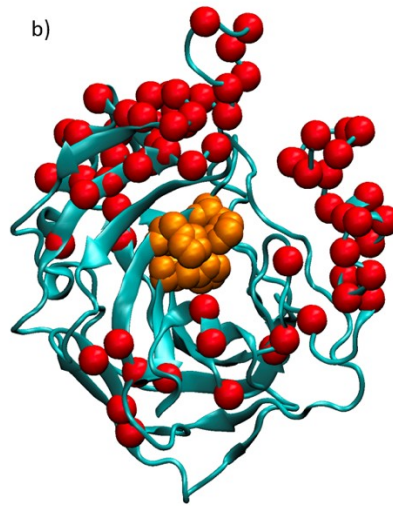 c) 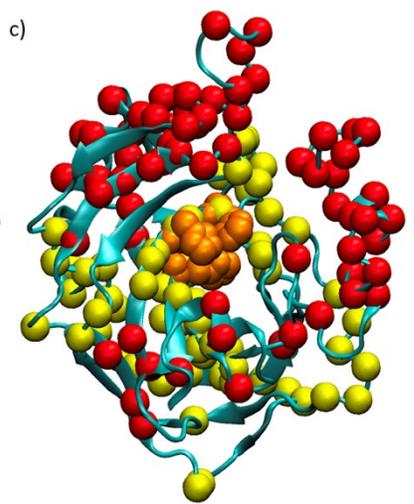

**Figure 3**. Cartoon representation of enzymatic systems for allocation of the best/worst donors upon the three-dimensional structure with catalytic residues highlighted in orange as reference. Red (yellow) spheres represent the best (worst) entropy and information donors. I) Human Lysozyme, a) catalytic residues Glu35 and Asp 53, b) addition of the 50 best and c) 50 worst donors. II) Ribonuclease A1, a) catalytic residues HIS 12, Lys 41 and His 119, b) addition of the 50 best and c) 50 worst donors. III) Pepsin, a) catalytic residues Asp32 and Asp215, b) addition of the 100 best and c) 100 worst donors. IV) β-lactamase, a) catalytic residues Lys 73, Ser 70, Ser 130, Asn 132, Glu 166 and Lys 234 b) addition of the 100 best and c) 100 worst donors. V) Human glucokinase, a) catalytic residues Thr 168, Lys 169 and Asp 205, b) addition of the 150 best and c) 150 worst donors. VI) Carbonic anhydrase II, a) catalytic residues His94, His96 and His 119, b) addition of the 70 best and c) 70 worst donors.

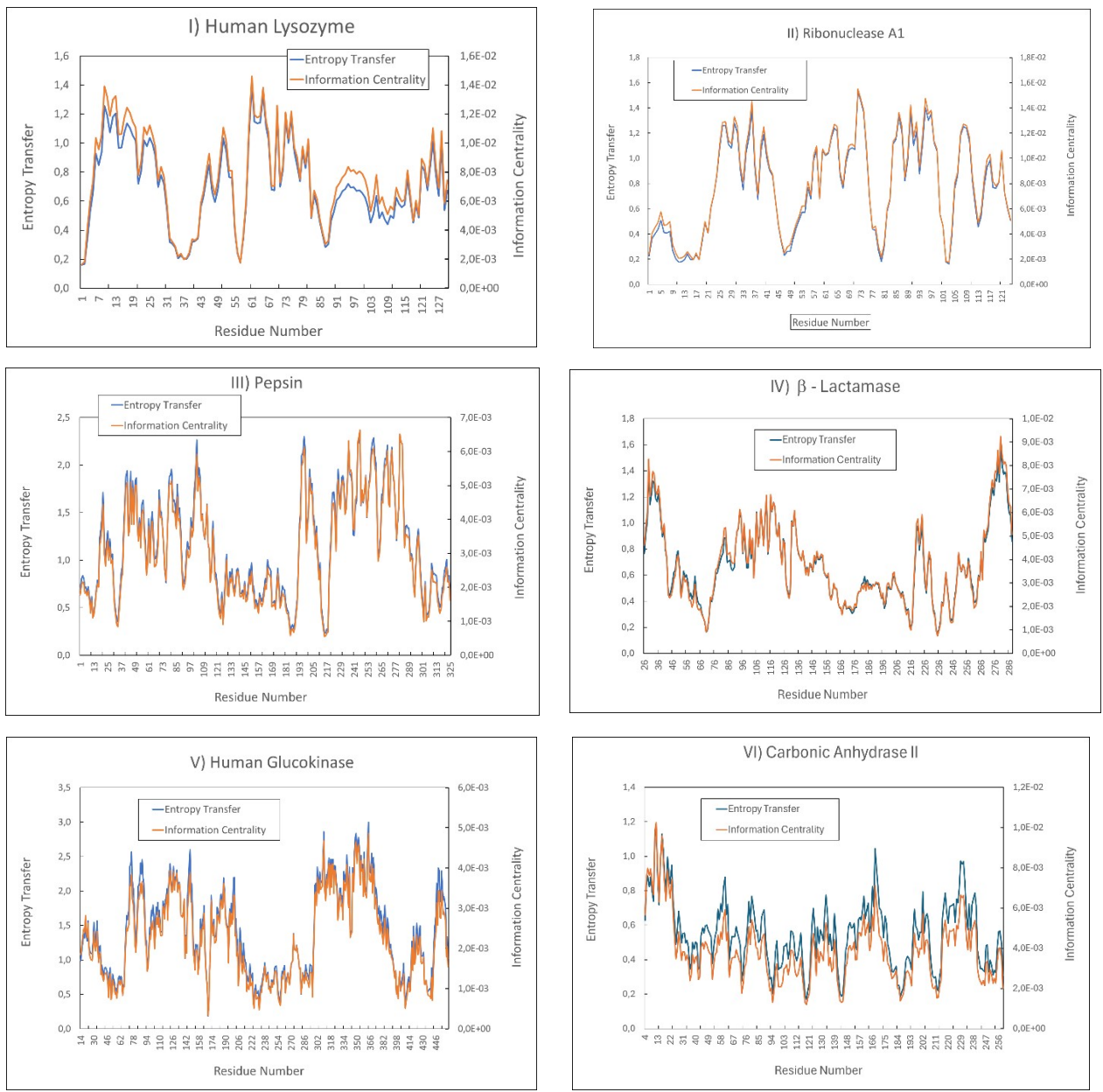

**Figure 4**. Comparison between the entropy transfer and information centrality outputs. I) Human Lysozyme. II) Ribonuclease A1. III) Pepsin. IV) β-lactamase. V) Human glucokinase. VI) Carbonic anhydrase II. Three last cases do not start at residue number one (in the abscissa) due to missing residues in the PDB structures.

## Discussion and Conclusion.

The modulation of enzyme activity is central for the fine control that maintain the metabolism in living beings. In this article we make a general analysis of the entropy and information flow throughout the structure of seven different enzymatic systems, whether they present or not allosteric regulation. We found a generalized feature of entropy and information flow in enzymes, that is, there is a gradient of flow from the peripherical sites towards the catalytic sites of them. This is a general trend in all the analyzed systems. The variability and differences among the analyzed systems suggest that this is a general feature of all enzymes, in which information is transferred to the catalytic site. We reach this result by using two different algorithms, dGNM and IC. Both methods share a common root, thai is, the Kirchhoff matrix $\Gamma$ or the graph Laplacian and it pseudo inverse $\Gamma^{-1}$. Both methodologies gives the same results with minor differences. While dGNM offers a dynamical perspective, the IC method offers a topological reading of this phenomenon. The IC identifies the same residues as the dGNM entropy transfer without requiring time-delayed correlations. This means that the ensemble of allosteric pathways is not an emergent property of the dynamics alone, but is already encoded in the contact topology of the protein: residues with high IC are structurally positioned as hubs in the network, and it is this topological privileged position that makes them preferential conduits for information flow. This distinction has practical consequences. Erman's model emphasizes that entropy alone is insufficient to determine allosteric communication, and that causality requires the introduction of time-delayed correlations (Hacisuleyman 2017). That is correct at the level of individual residue-to-residue directionality. However, the agreement between $T_{i \to \varnothing}(\tau)$ (equation 5) and and $IC_i$ (equation 9) shown in Figure 4 suggests that, when one is interested in identifying the entropy or the information flow directionality, the structural and topological analysis is more than sufficient. The Laplacian of the contact graph already contains the information needed to locate the entropy sources and sinks. This reinforces the value of graph-theoretical and topological approaches to the study the allosteric features of protein and enzyme molecules. Thus, the three-dimensional fold and its connectivity determine where allosteric signals converge. Further research should next be performed to establish the details of this allosteric entropy and information flow as consequence of conformational changes, and other relevant chemical events, during the catalytic cycle of some to the selected analyzed systems.

**Statement.** During the preparation of this work no generative AI was used.

**Acknowledgements.** J.M. Gonzalez and G. Miño-Galaz are thankful UNAB for a grant DI-17-20/REG-VRID. Powered@NLHPC: This research/thesis was partially supported by the supercomputing infrastructure of the NLHPC (CCSS210001)